# The Multiple viewpoints as approach to information retrieval within Collaborative Development Context

Hichem M. Geryville, Yacine Ouzrout, Abdelaziz Bouras, and Nikolaos S. Sapidis

*Abstract*—Nowadays, to achieve competitive advantage, the industrial companies are considering that success is sustained to great product development. That is to manage the product throughout its entire lifecycle. Achieving this goal requires a tight collaboration between actors from a wide variety of domains, using different software tools producing various product data types and formats. The actors' collaboration is mainly based on the exchange /share product information. The representation of the actors' viewpoints is the underlying requirement of the collaborative product development. The multiple viewpoints approach was designed to provide an organizational framework following the actors' perspectives in the collaboration, and their relationships. The approach acknowledges the inevitability of multiple integration of product information as different views, promotes gathering of actors' interest, and encourages retrieved adequate information while providing support for integration through PLM and/or SCM collaboration. In this paper, a multiple viewpoints representation is proposed. The product, process, organization information models are discussed. A series of issues referring to the viewpoints representation are discussed in detail. Based on XML standard, taking cyclone vessel as an example, an application case of part of product information modeling is stated.

## I. INTRODUCTION

COLLABORATIVE product development constitutes the core activity of many industrial companies. It includes actually a variety of business processes associated with the activities appearing in the product lifecycle. This is where PLM (Product Lifecycle Management) comes in. PLM system is the solution to support management of product information definition throughout its complete lifecycle. However, this system has not good filter methods for pertinent information extraction, especially to exchange and share information between actors. Also, Several different interpretations of the same information often exist, and many metadata contain information that is related or complementary to other metadata, but these different interpretations are rarely integrated, and often are only used together in an ad hoc way; so that an actor must keep switching back and forth between different applications and interfaces in the course of a single seek. In this context, the multiple viewpoints system is key element to integrate more than one interpretation approach across a body of information, gaining leverage in retrieval from their diversity and simplifying actor interactions by uniting them under a common interface. Where the multiple viewpoints approach permits to capture/retrieve the adequate information following the actors' interests/objectives on the collaboration, based on his own experiences/knowledge.

The remainder of this paper is organized as follows: section 2 gives an overview of the main viewpoints approach. Section 3 is dedicated to the description of our multiple viewpoints system. In section 4, an example of product is studied and used for instantiating the viewpoints approach. The viewpoint instantiation is then used to show how an effective information extraction is achieved. The paper is ended with some concluding remarks and important perspectives in section 5.

## II. LITERATURE REVIEW

We use the term viewpoint to mean a scheme for representing a collection of information objects, along with a mechanism for accessing this information. The term has also been used, with some frequency and great inconsistency, in the areas of information visualization and Human Computer Interfaces (HCI). Researchers such as Teraoka and Maruyama [1] are generally interested in representing a user's interests and purpose. They use "multiple viewpoints" to parameter an information visualization system that indicates how to present information based on a particular interest profile. While a user's interests might serve as the basis for a viewpoint, the viewpoints are not limited to different visualizations of the same information relationships; the relationships themselves may differ as well.

Ribière [2] considers "Viewpoint" as a polysemous word,

Manuscript received January 30, 2007.
H. Geryville, Y. Ouzrout and A. Bouras are with the CERRAL/LIESP laboratory, Lumiere University of Lyon, 160 Boulevard de l'Université, 69676 Bron Cedex, Lyon, FRANCE (phone: +33478772670; fax: +33478006328; e-mail: hichem.geryville@univ-lyon2.fr, yacine.ouzrout@univ-lyon2.fr, abdelaziz.bouras@univ-lyon2.fr).
N. Sapidis is with Department of Product and Systems Design Engineering, University of The Aegean, 84100 Syros, GREECE (e-mail: sapidis@aegean.gr).

i.e. its definition depends on the context of use. She defines a viewpoint as "*a perspective of interest from which an expert examines the knowledge base*". It is a general definition that can take several interpretations in different domains of application. She proposes an extension of the conceptual graph formalism to integrate viewpoints in the support and in the building of conceptual graphs. Where the viewpoints allow her to define the context of use and the origin actor of concept types introduced in a graph. The aim of her proposal is to define viewpoints to help knowledge representation with conceptual graphs for multi-expert knowledge acquisition and also to have an accessible and evolutive knowledge base of conceptual graphs through viewpoints.

Powell and French [3] introduce the concept of multiple viewpoints by using both multiple relevance judgments and multiple representations together. Collection selection [4] may also prove an important element in the design of multiple viewpoint algorithms that attempt to gain efficiency by consulting only a subset of the viewpoints in a system.

In the collaborative development, a lot of studies focusing on knowledge representation and product modelling during collaborative development process have been investigated [5], [6], [7]. However, most of them emphatically address only a part of the entire issue of product modelling. And, up until now, few papers have synthetically investigated product information modelling based on the users' points of view on a large-system, from the design of product information modelling until there use in different application.

Geryville and all. [8] describe the concept of the multiple viewpoints as a composition of four components: the actor, the viewpoint's domain, its related objectives (objectives of the actors in the collaboration), and its relationships with the other viewpoints. This definition is characterized by a context, which allows the restitution of the information that the actor wants use/retrieve, and a degree of importance of each viewpoint.

The scope of the presented works mainly deals with the integration of an information visualization interface, where the same information has different meaning following its relationships. However their limitation is that none of them integrate the real interests of the actors within collaborative product context, such as the extraction/retrieve of adequate information. In multidisciplinary collaboration, the actors need to integrate their viewpoints on the product along its lifecycle stages, where they also need to retrieve important information within different stages according to their interests.

Based on previous studies of collaborative product development process [8],[9], this paper attempts to contribute to give a complete definition for actors' viewpoints representation, where we propose an approach that makes connections with product, process, and organization information for a complete interaction between multidisciplinary actors. To give an appropriate definition of multiple viewpoints actors, we situate the actors under different views (product, process and supply chain organization). In the next section, we define our concepts of multiple viewpoints and its interactions with the product/process information within a collaborative supply chain organization.

### III. PROPOSED ARCHITECTURE

In multidisciplinary collaboration, the framework must be characterized by the following features:
-- a base-level information model should contain enough information for various needs of the product collaborators, i.e., the product information model should be based on the whole lifecycle of the product to assure it contains complete information meeting the various demands of all actors in all product development stages.
-- actors' viewpoints should represent the actors' knowledge and interests on the product collaboration which permit to help them to seek, extract, exchange the adequate information. In a word, a whole-life-cycle product information model with viewpoint representation is necessary to support collaborative product development.

#### A. The Information Model

In a previous studies [9], [10], we defined an information model called PPCO (Product–Process–Collaboration–Organization information model) based, especially, on the collaborative and project options, this study is inspired 50% from Gzara's work [11] where they propose the Product–Process–Organization model (PPO). The PPCO provides a base-level information model that is open, extensible, independent from any product development process; and aiming at capturing the engineering and business context commonly shared in product development.

The PPCO model is based on the four elements stated previously: product, process, collaboration and organization.

1) **Product:** the architecture of the product is defined not only by the decomposition of the final product into components, functions, behaviours, etc, but also by the interactions between all these components. The product metadata model uses the Core Product Model [13].

2) **Process:** the product development process is generally a complex procedure involving information exchange across the many activities/tasks in order to execute the collaborative work. Various network-based methods have been used to map and study development processes.

3) **Organization:** the organization structure determines who works with whom and who reports to whom. However, in supply chains organization we are particularly interested in the study of the communication patterns between the actors conducting the technical development work.

#### B. The multiple Viewpoints Approach

A viewpoint is any structure from which we can elicit an informative result from a collection of information by presenting a query; which is based on some set of relationships among the data. In this section, we define the primary elements of our framework for describing viewpoints and multiple viewpoints approach:

*1) Single Viewpoints*

There are myriad methods of representing a single information collection and the relationships among its elements, including information retrieval models, graphs, and hierarchical arrangements. Each viewpoint in a multiple viewpoint system contains a representation of some set of information items, a subset of the universe of information items with which the system is concerned. This is the viewpoint domain.

We are less concerned with the representational schema used in a viewpoint and more interested in the interface to the viewpoint; this is the viewpoint access method. Primarily, we need to know the query language, the language of queries that are valid input to the viewpoint as a function that maps elements of the viewpoint query language to results, subsets of the viewpoint domain.

So, we consider that a viewpoint implements the condition for an *Actor* to interpret the sense of a collection of information: it is defined by the *Object*, on which the interpretation is performed, the Actor performing it, the *Expression* and *Content* of the interpretation of the Object by the Actor, and the *Context* in which this interpretation is performed.

*2) Approach of Multiple Viewpoints*

A system of multiple viewpoints consists of: some set of viewpoints, which represents relationships among the elements of some subset of the universe of artefacts which the concern system; a set of transition mappings for pairs of viewpoints in the system; and a merge function that defines how to construct a result set based on a set of viewpoint result sets. An initial query to the multiple systems might feel some translation and increasing to be usable as a generic query for any viewpoint.

*3) Transition Mappings*

The transition mapping is at the heart of the concept of multiple viewpoints framework; although the transition mapping does not determine when or to which viewpoint we should change, it implicitly describes the relationship between two viewpoints by providing corresponding entrance points for each, and thereby helps us to maintain the context of the seek when switching viewpoints.

Let's suppose a simple system with two viewpoints. If we have consulted one of these with a query and we are unsatisfied with the result, how should we move to the other viewpoint? The simplest method is to consult the second viewpoint with the same initial query; this is appropriate if we have information about the relationship between the two viewpoints.

A transition mapping from one viewpoint to another must transform elements of the query language of the first viewpoint to elements of the query language of the second. If every element of the each viewpoint domain is a valid viewpoint query, we can always perform a transition based on elements in the intersection of two viewpoint domains; that is, we can move from the representation of certain information in the first viewpoint to the representation of the same information in the second.

A transition mapping from one query to another may take advantage of known synonyms in two indexing vocabularies, or might make changes based on more complex relations.

In some cases, it will be reasonable to create transition mappings wholly or partly by actors, based on human judgments. In other cases where we have sufficient information about two viewpoint representations, data-mining techniques may be suitable for more automatic discovery of appropriate mappings.

*4) Merging Results*

For a system of viewpoints to usually present the results of a series of viewpoint consultations, a well-designed merge method is essential. The merge method must take into account not only any rankings of results from a particular viewpoint, but also any known differences in reliability among viewpoints. The merging algorithm must also handle the case where the same element of the information system appears in two different viewpoint result sets. In many cases, it will be convenient for the system result to appear as a single ranked list, but such formatting must depend on the intended purpose of the system; this is primarily an HCI issue.

The job of the merge function becomes quite complex when the query has been changed from its initial state. Possibly, results should be ordered based on the similarity of the query that engendered them to the original query, but determining a useful measure of similarity is nontrivial. Then again, it may be that revealing this information to the user will only make the result confusing.

## IV. EXAMPLE

To illustrate the proposed approach, we present an example about a closed pack cyclone vessel. In the next sections, we present the three models, product structure (or decomposition), process development, and the development of the supply-chain organization. In [8] you can find more details on this example.

### A. Product Structure Example

The closed cyclone structure is composed of 18 components. The product architecture and the interactions between the components are documented and detailed in this model (not presented in this paper.

### B. Development Process Example

The process tree describes manufacturing process to determine the feasible layout of the customer requirements of the closed pack cyclone vessel. This is based on a digital model using CAD solid models. Interactions in this type of model represent flows of information and data between the tasks of the activities (in this example, we consider that every process has only one activity).

*C. Supply Chain Organization Example*

The supply chain organization defines the enterprises integrated to develop a new cyclone vessel project. The organization involves 3 teams; each team has a responsibility for a major component. The given matrix depicts the interactions across the 3 teams in terms of frequency of their required collaboration.

*D. Information Restitution by Viewpoints Approach*

Let's take the example of an actor "ActorX" integrating the supply chain organization into the team 1 as *an external designer*. He has two focuses on the product "cyclone vessel", the first as **shape global design**, and the second as **mechanical design** (Fig. 1). The actor's objectives are related respectively to the activity on geometry and mechanic tasks. To retrieve the adequate information for the actor "ActorX", we need to filter and classify the information following his interests. By using the adequate query, the framework compares the information of each viewpoint and gives the sets of adequate information to the actor following his focuses on the product and his activities in the project. Based on the level's batch definition, the system regroups the batches with high-level hierarchy, and retrieves the information which is more adequate to the actor according to his focuses and activities [8].

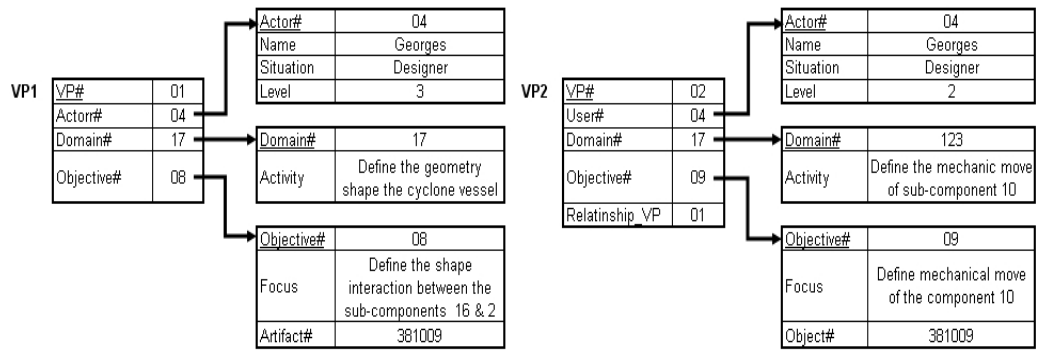

Figure 1. ActorX Viempoint

## V. Conclusion

Multidisciplinary collaboration (MC) is a complex domain, in which all the actors need to exchange and share product and process information. In fact, product information generated by each actor is communicated to all actors in order to integrate them in a shared representation. Knowledge about actors' preferences, methods, etc. and about actors' focuses, constraints, objectives, etc. must be taken into account to manage information extraction.

In MC, the use of viewpoint in the structured collaborative product development shows how the viewpoint notion can provide real help in the extraction, treatment and consulting of adequate product/process information. The proposed viewpoint description and multi-level management approach aim to structure the actor's focuses in multidisciplinary collaboration thanks to a more accurate characterization of the viewpoints, which allows an intelligent indexation of the product/process information.

Actually, we implement the multiple viewpoints framework on our prototype, which integrate the PPCO model, and we will test it with 2 developed scenarios (PLM and SCM).